\documentclass[12pt]{article}

\usepackage{latexsym}

\begin{document}

\title{The physical graviton
two-point function in de Sitter spacetime with $S^{3}$ spatial sections }

\author{Atsushi Higuchi$^1$ and Richard H.\ Weeks$^2$\\
{\normalsize Department of Mathematics, University of York}\\
{\normalsize Heslington, York, YO10 5DD, United Kingdom}\\
{\normalsize $^1$Email: ah28@york.ac.uk}\\
{\normalsize $^2$Email: rhw101@york.ac.uk}}

\date{9 June, 2003}
\maketitle

\begin{abstract}
We compute the physical graviton two-point function in de Sitter spacetime
with three-sphere spatial sections.  We demonstrate that the large-distance 
growth present in the corresponding two-point function 
in spatially flat de~Sitter spacetime is absent.
We verify that our two-point function agrees with that in Minkowski spacetime
in the zero cosmological constant limit. 
\end{abstract}

\renewcommand{\theequation}{\arabic{section}.\arabic{equation}}

\section{Introduction}

It is known that the graviton two-point functions in de~Sitter spacetime 
in various coordinates and gauges~\cite{AllenTuryn,Flo,Anto,Tsam} 
grow as functions of the
coordinate distance.  For example, the physical two-point function with
two polarizations grows logarithmically at large distances in
spatially flat de Sitter spacetime~\cite{Allen87:1}. 
(The term ``physical" here is used
to indicate 
that the gauge degrees of freedom are completely fixed.) 
If this large-distance growth
of graviton two-point functions manifested itself in physical quantities,  
it could have serious implications in inflationary 
cosmologies~\cite{Infl1,Infl2,Infl3}.  
Thus, it is interesting to determine whether
this growth depends on the gauge choice.  This clarification may also be 
helpful in understanding the
mechanism for damping the cosmological constant~\cite{Tsamis} proposed
a few years ago.
(The analogous two-point function
in hyperbolic de~Sitter spacetime does not grow at large coordinate 
distances~\cite{Hawketal}.  However, the hyperbolic coordinate system
does not cover the region where the two-point function grows either in
the covariant gauge~\cite{AllenTuryn} or in the physical 
gauge~\cite{Allen87:1}.)

It was shown recently
that one can gauge away the large-distance growth in the physical gauge
in spatially flat coordinate system~\cite{gaugeartefact} and
in the covariant gauge~\cite{HK2}.  In this paper we calculate the
physical two-point function
in de~Sitter spacetime with three-sphere ($S^3$) spatial sections
and demonstrate that it remains finite (as long as one of the point
is fixed) in the
entire spacetime except for the light-cone singularity.  This 
shows decisively that the large-distance growth in other gauges is
a gauge artefact in the following sense: it does not cause similar growth
in the two-point function of any
locally-constructed gauge-invariant operator, such as the linearized Weyl
tensor~\cite{Kouris}.\footnote{If we allow both points to go to infinity,
the graviton
two-point function we obtain
can grow logarithmically.  This fact does not invalidate
our assertion here because one point can always be moved to the origin
when the two-point function of a locally-constructed gauge-invariant  
operator is evaluated.}  (This fact might be related to the 
observation that there is no classical instability due to gravitational
clumping of thermally excited
gravitons in de~Sitter spacetime~\cite{GP}.) 
The rest of the paper is organized as follows. 
We summarize basic facts about linearized gravity in de~Sitter spacetime
which are relevant to this paper in section 2.
Then we present the details of our 
calculation of the two-point function in section 3 and discuss some of its
properties in section 4.  We present the calculation of the corresponding
two-point function in Minkowski spacetime in Appendix A and explain the
relationship between the coordinate system we use here and other coordinate
systems for de~Sitter spacetime.  
We adopt the metric signature
$(-+++)$ throughout this paper.  We have used the computing package Maple
7 in many of our calculations.

\section{Linearized gravity in de~Sitter spacetime}
\setcounter{equation}{0}

We need the explicit form of the transverse-traceless tensor
spherical harmonics on $S^3$.
They can be found, e.g. in Ref.\ \cite{Tensor}, which we closely 
follow here.
The metric on the unit $S^3$ is given by 
\begin{equation}
ds^{2} = d\chi^{2} +\sin^{2}\chi (d\theta^{2} + \sin^{2}\theta
d\varphi^{2})\,,
\end{equation}
where $0\leq \chi \leq \pi$, $0\leq \theta\leq \pi$ and
$0 \leq \varphi \leq 2\pi$.  Let $\eta_{ab}$ and $D_c$ denote the metric
and the covariant derivative operator, respectively, on $S^3$.
The transverse-traceless (TT) tensor spherical harmonics $T_{ab}^{(C;Llm)}$
are symmetric tensors satisfying
$D^b T_{ab}^{(C;Llm)} = 0$, $\eta^{ab}T_{ab}^{(C;Llm)} = 0$ and
\begin{equation}
D_c D^c T_{ab}^{(C;Llm)} = [-L(L+2) + 2]T_{ab}^{(C;Llm)}\,,
\end{equation}
where $L\,(\geq 2)$ is an integer.
The label $C$ is either v (vector) or
s (scalar).  The labels $l$ and $m$ are integers satisfying
$2 \leq l \leq L$ and $-l \leq m \leq l$. 
We also require the following normalization condition:
\begin{equation}
\int d\Omega\,\overline{T^{(C;Llm)}_{ab}} T^{(C';L'l'm')ab} =
\delta_{CC'}\delta_{LL'}\delta_{ll'}\delta_{mm'}\,,
\end{equation}
where $d\Omega = d\chi\,\sin^2\chi\, d\theta\,\sin\theta\,d\varphi$ is
the volume elment of $S^3$.  
We define the following functions:
\begin{equation}
\bar{P}_L^l(\chi) \equiv \left[ \frac{(L+1)(L+l+1)!}{(L-l)!}\right]^{1/2}
(\sin\chi)^{-1/2}{\rm P}_{L+1/2}^{-(l+1/2)}(\cos\chi)\,,
\label{barP}
\end{equation}
where the Legendre functions ${\rm P}_\nu^{-\mu}(x)$ are given in terms
of the hypergeometric functions as~\cite{Grad}
\begin{equation}
{\rm P}_\nu^{-\mu}(x) \equiv
\frac{1}{\Gamma(1+\mu)}\left( \frac{1-x}{1+x}\right)^{\mu/2}
F\left[ -\nu-1,\nu;1+\mu;(1-x)/2\right]\,.
\end{equation}
We also define the transverse vector spherical harmonics 
on the unit two-sphere ($S^2$) as
\begin{equation}
V_i^{(lm)}(\theta,\varphi) \equiv \frac{1}{\sqrt{l(l+1)}}
\epsilon_{ij}\partial^j Y_{lm}(\theta,\varphi)\,,
\end{equation}
where $\epsilon_{ij}$ is the invariant antisymmetric tensor with
$\epsilon_{\theta\varphi} = \sin\theta$. The indices $i,j = \theta,\varphi$ 
are raised and lowered by the metric $\tilde{\eta}_{ij}$ on the unit $S^2$.
Then the TT tensor spherical harmonics
are given as follows: 
\begin{eqnarray}
T_{\chi\chi}^{({\rm v};Llm)} & = & 0\,, \\
T_{\chi i}^{({\rm v};Llm)} & = & 
\left[\frac{(l-1)(l+2)}{2L(L+2)}\right]^{\frac{1}{2}} \bar{P}_{L}^{l}(\chi
)V_{i}^{(lm)}(\theta ,\varphi)\,, \label{vchii} \\
T_{ij}^{({\rm v};Llm)} & = & 
\frac{\sin^{2}\chi}{[2(l-1)(l+2)L(L+2)]^\frac{1}{2}}\left(\frac{\partial}
{\partial\chi} +
2\cot\chi\right )\bar{P}_{L}^{l}(\chi )[\tilde{D}_i V_{j}^{(lm)}(\theta 
,\varphi)
+(i\leftrightarrow j)],\nonumber \\
\end{eqnarray}
with $\tilde{D}_i$ denoting the covariant derivative on the unit $S^2$, and
\begin{eqnarray}
T_{\chi\chi}^{({\rm s};Llm)} 
& = & \frac{c_{L}^{l}}{\sin^{2}\chi}\bar{P}_{L}^{l}
(\chi)Y_{lm}(\theta ,\varphi)\,, \label{Tchi}\\
T_{\chi i}^{({\rm s};Llm)} & = & 
\frac{c_{L}^{l}}{l(l+1)}\left(\frac{\partial}{\partial\chi} +
\cot\chi\right)\bar{P}_{L}^{l}(\chi )\tilde{D}_{i}Y_{lm}(\theta ,\varphi )\,,\\
T_{ij}^{({\rm s};Llm)} & = & 
c_{L}^{l}\left\{ F_L^l(\chi )\left[\tilde{D}_{i}\tilde{D}_{j} +
\frac{l(l+1)}{2}\tilde{\eta}_{ij}\right]Y_{lm}(\theta ,\varphi ) -
\frac{\tilde{\eta}_{ij}}{2}
\bar{P}_{L}^{l}(\chi )Y_{lm}(\theta ,\varphi )\right\}\,. \nonumber \\
\label{misprint}
\end{eqnarray}
(There is a misprint in the equation corresponding to (\ref{misprint}) 
in Ref.\ \cite{Tensor}.\footnote{ 
We thank R.\ Caldwell for pointing this out.})
The function $F_L^l(\chi)$ and the constant $c_{L}^{l}$
are defined as follows:
\begin{eqnarray}
F_L^l(\chi ) & = & 
\frac{2\sin^{2}\chi}{(l-1)l(l+1)(l+2)}\left(\frac{d\ }{d\chi} +
2\cot\chi\right )\left(\frac{d\ }{d\chi} + \cot\chi\right) 
\bar{P}_{L}^{l}(\chi ) \nonumber \\
&& - \frac{1}{(l-1)(l+2)}\bar{P}_{L}^{l}(\chi)\,, \\
c_{L}^{l} & = & 
\left[\frac{(l-1)l(l+1)(l+2)}{2L(L+1)^{2}(L+2)}\right]^{\frac{1}{2}}\,.
\label{CLL}
\end{eqnarray}

The metric of de~Sitter spacetime with $S^3$ spatial sections is given by
\begin{equation}
ds^2 = H^{-2}\left\{
-dt^2 + \cosh^{2}t[d\chi^{2} + \sin^{2}\chi (d\theta^{2} +
\sin^{2}\theta d\varphi^{2})]\right\}\,. \label{back}
\end{equation}
This is a solution to Einstein's equations with 
cosmological constant $\Lambda = 3H^2$: 
\begin{equation}
R_{\mu\nu} -\frac{1}{2}g_{\mu\nu}R+3H^2 g_{\mu\nu}=0\,.
\end{equation}
The linearized field equation is obtained by setting $g_{\mu\nu}
=\hat{g}_{\mu\nu} + h_{\mu\nu}$, where $\hat{g}_{\mu\nu}$  is the background 
metric (\ref{back}). We find
\begin{eqnarray}
&& \frac{1}{2}\Box h_{\mu\nu} - \frac{1}{2}(\nabla_{\mu}\nabla_{\alpha}
h^{\alpha}_{\nu} - \nabla_{\nu}\nabla_{\alpha}
h^{\alpha}_{\mu}) + \frac{1}{2}\nabla_{\mu}\nabla_{\nu}h\nonumber \\
&& - \frac{1}{2}\hat{g}_{\mu\nu}\Box h +
\frac{1}{2}\hat{g}_{\mu\nu}\nabla_{\alpha}\nabla_{\beta}h^{\alpha\beta} - 
H^2(h_{\mu\nu}
+ \frac{1}{2}\hat{g}_{\mu\nu}
h) = 0\,. \label{feq}
\end{eqnarray}
Here, $\nabla_{\mu}$ is the background covariant derivative,
indices are raised and lowered by the background metric, and $h$ is the
trace of $h_{\mu\nu}$ with respect to the background metric. 
We have also defined 
$\Box \equiv \nabla_\alpha\nabla^\alpha$.  We will denote the
background metric $\hat{g}_{\mu\nu}$ simply by $g_{\mu\nu}$ from now on.

The field equation (\ref{feq}) is invariant under the gauge transformation
\begin{equation}
\delta h_{\mu\nu} = \nabla_{\mu}\Lambda_{\nu} +
\nabla_{\nu}\Lambda_{\mu}.
\end{equation}
It is known 
that one can impose the conditions $h_{0\mu} = 0$,
$D_a h^{ab} = 0$ and $\eta^{ab}h_{ab} = 0$, with $a,b = \chi,\theta,\varphi$
and that these conditions fix the
gauge degrees of freedom completely (see, e.g. Ref.\ \cite{Tensor}). 
Then the field $h_{\mu\nu}$ satisfies
\begin{equation}
(\Box - 2H^2)h_{\mu\nu} = 0\,.
\end{equation}
Independent solutions to this equation satisfying the gauge conditions 
are~\cite{Tensor}
\begin{eqnarray}
h_{ab}^{(C;Llm)}(t,\chi,\theta,\varphi) & = & 
\left[\frac{L+1}{L(L+2)}\right]^{1/2}\cosh t\left(1 + \frac{i\sinh t}
{L+1}\right)\left(\frac{1 - i\sinh t}
{1+i\sinh t}\right)^{(L+1)/2}\nonumber \\
&& \times T_{ab}^{(C;Llm)}(\chi ,\theta ,\varphi)\,.
\label{sol1}
\end{eqnarray}
These are the coefficient functions of the annihilation operators for the
standard vacuum state known as the Euclidean~\cite{GibHawk} or
Bunch-Davies~\cite{BD} vacuum. 
They satisfy the following normalization condition:
\begin{equation}
\langle h^{(C;Llm)}| h^{(C';L'l'm')}\rangle = 
H^2\delta_{CC'}\delta_{LL'}\delta_{ll'}\delta_{mm'}\,,
\end{equation}
where we have defined
\begin{eqnarray}
\langle h^{(1)}\left| \right.  h^{(2)}\rangle & \equiv &
\frac{iH}{2}\int
d\Sigma \left[ \overline{h^{(1)\mu\nu}}\nabla_{0}h^{(2)}_{\mu\nu}
- h^{(2)\mu\nu}\nabla_{0}\overline{h^{(1)}_{\mu\nu}}\right]\nonumber \\
& = & \frac{iH^2}{2\cosh t}\int
d\Omega \left[\overline{h^{(1)ab}}\dot{h}^{(2)}_{ab} - 
h^{(2)ab}\overline{\dot{h}^{(1)}_{ab}}\right]\,,
\end{eqnarray}
with $d\Sigma \equiv H^{-3}\cosh^3 t\,d\Omega$ being the volume element of the
$S^3$ Cauchy surface.  In the first line of this equation the indices
$\mu$ and $\nu$ are raised and lowered by the de~Sitter metric $g_{\mu\nu}$
whereas in the second line the indices $a$ and $b$ are raised and lowered
by the metric $\eta_{ab}$ on the unit $S^3$.

Let us introduce a new time variable $\tau$ satisfying
$-\pi/2 < \tau < \pi/2$ by
\begin{equation}
\tan\tau \equiv \sinh t\,.
\end{equation}
Then the metric (\ref{back}) becomes
\begin{equation}
ds^2 = H^{-2}\sec^2\tau [-d\tau^2 + d\chi^2 + 
\sin^2\chi(d\theta^2 + \sin^2\theta d\varphi^2)]\,.
\end{equation}
The solutions (\ref{sol1}) become
\begin{equation}
h_{ab}^{(C;Llm)} =
\frac{i\sec\tau}{\sqrt{L(L+1)(L+2)}}D(\tau)e^{-i(L+1)\tau}
T_{ab}^{(C;Llm)}(\chi ,\theta ,\varphi)\,, \label{sol2}
\end{equation}
where
\begin{equation}
D(\tau) \equiv \frac{\partial\ }{\partial \tau} + \tan\tau\,.
\label{Dtau}
\end{equation}

The graviton two-point function is then given by
\begin{equation}
G_{aba'b'}(x,x')  =  
\frac{16\pi G}{H^2} \sum_{C}\sum_{L=2}^{\infty}
\sum_{l=2}^{L}\sum_{m=-l}^{l}h_{ab}^{(C;Llm)}(x)
\overline{h^{(C;Llm)}_{a'b'}(x')}\,,
\end{equation}
where $x\equiv (\tau,\chi,\theta,\varphi)$ and 
$x'\equiv (\tau',\chi',\theta',\varphi')$ and where $G$ is Newton's constant.  
(The factor of $16\pi G$ arises due to the factor $(16\pi G)^{-1}$ in the
Einstein-Hilbert action.) 
By substituting (\ref{sol2}) we obtain
\begin{eqnarray}
\hat{G}_{aba'b'}(x,x') & \equiv & 
(16\pi GH^2)^{-1} G_{aba'b'}(x,x')\nonumber \\
& = &   
H^{-4}\sec\tau\sec\tau'D(\tau)D(\tau')\nonumber \\
&& \times \sum_{C}\sum_{L=2}^\infty \sum_{l=2}^L
\sum_{m=-l}^l \frac{e^{-i(L+1)(\tau-\tau')}}{L(L+1)(L+2)}
T_{ab}^{(C;Llm)}(\chi,\theta,\varphi)
\overline{T_{a'b'}^{(C;Llm)}(\chi',\theta',\varphi')}\,.\nonumber \\
\label{twoP}
\end{eqnarray}

\section{The two-point function}

\setcounter{equation}{0}

In order to write down the explicit form of the two point function 
(\ref{twoP}) we need to introduce some definitions related to the (shorter) 
geodesic 
on the unit $S^3$ connecting the points $p\equiv (\chi, \theta, \varphi)$ and
$p'\equiv (\chi',\theta',\varphi')$.  They will be analogous to those 
introduced
in Ref.\ \cite{Allen87:1} for the spatially flat case.  We assume that
the point $p'$ is not the antipodal point of
$p$ so that there is a unique shortest geodesic connecting them.\footnote{
There are two geodesics connecting two generic points on $S^3$.  By the
geodesic we mean the shorter one in the rest of this paper.}  

Let $n^{a}$ and $n^{a'}$ be the unit tangent vectors to the geodesic at
points $p$ and $p'$, respectively, which are pointing away 
from each other.  We define
the vectors $N^{a}$ and $N^{a'}$ at points 
$x \equiv (\tau,p)$ and $x' \equiv (\tau',p')$, respectively,
in de~Sitter spacetime as
\begin{eqnarray}
N^{a} & \equiv  & H\cos\tau\,n^a\,,\\ 
N^{a'} & \equiv &  H\cos\tau'\,n^{a'}\,.
\end{eqnarray}
These vectors can be interpreted as unit vectors in de Sitter spacetime.
We define $g_{ab}$ ($g_{a'b'}$)
to be the symmetric tensor on the unit $S^3$ obtained as the
space components of the metric tensor $g_{\mu\nu}$ at $x$ ($x'$).  Thus,
$g_{ab} = H^{-2}\sec^2\tau\,\eta_{ab}$ and 
$g_{a'b'} = H^{-2}\sec^2\tau'\,\eta_{a'b'}$.
Next we define the parallel propagator ${P^{a'}}_{a}$ on the unit $S^3$ as 
follows.  Let $V^{a}$ be a tangent vector at $p$.  We let 
${P^{a'}}_{a} V^{a}$ be the vector obtained by parallelly
transporting $V^{a}$ along the geodesic from $p$ to $p'$.  We define
${P^a}_{a'}$ in a similar manner.  One has $P_{a'a} = P_{aa'}$. 
Note that 
${P^{a'}}_a n^{a} = - n^{a'}$ and that ${P^a}_{a'}n^{a'}= - n^{a}$. We then
define the tensors $g_{aa'}$ on $S^3$ for any two points
$(\tau,p)$ and $(\tau',p')$ in de~Sitter spacetime as 
\begin{equation}
g_{aa'} = H^{-2}\sec\tau\sec\tau'\,P_{aa'}.
\end{equation}

Due to rotational symmetry 
the two-point function $\hat{G}_{aba'b'}(x,x')$ 
must be a linear combination of bi-tensors
constructed from $N_a$, $N_{a'}$, $g_{ab}$, $g_{a'b'}$ and $g_{aa'}$. It
satisfies $g^{ab}\hat{G}_{aba'b'} = g^{a'b'}\hat{G}_{aba'b'}  = 0$ 
and is symmetric
under $a\leftrightarrow b$ and $a'\leftrightarrow b'$.  This implies that
\begin{equation}
\hat{G}_{aba'b'}(x,x') = 
g^{(1)}T^{(1)}_{aba'b'} + g^{(2)}T^{(2)}_{aba'b'}
+ g^{(3)}T^{(3)}_{aba'b'}\,,  \label{Gab}
\end{equation}
where
\begin{eqnarray}
T_{aba'b'}^{(1)} & \equiv & \left(
N_{a}N_{b} - \frac{1}{3}g_{ab}\right)\left(N_{a'}N_{b'} - 
\frac{1}{3}g_{a'b'}\right)\,, \\
T_{aba'b'}^{(2)} & = & g_{aa'}g_{bb'} + g_{ab'}g_{ba'}
-\frac{2}{3}g_{ab}g_{a'b'}\,, \\
T_{aba'b'}^{(3)} & = & g_{aa'}N_{b}N_{b'} + g_{ab'}N_{a'}N_{b} +
g_{a'b}N_{b'}N_{a} + g_{bb'}N_{a}N_{a'} + 4N_{a}N_{a'}N_{b}N_{b'}\,,
\end{eqnarray}
and where the $g^{(i)}$, $i=1,2,3,$ are functions of $\tau$, $\tau'$ and
the geodesic distance between $p$ and $p'$ on the unit $S^3$.  Thus,
we only need to find the functions $g^{(i)}$. 

Now, let us describe how to calculate these functions by relating them
to components of $\hat{G}_{aba'b'}$.
By using rotational symmetry on $S^3$ we may let 
$(\theta,\varphi) = (\theta',\varphi')$ and then
$\chi'\to 0$ without loss of generality.  Then, the geodesic between the
two points is along the line $(\theta,\varphi)=\,{\rm const}$ 
and the geodesic distance on the unit $S^3$ is
$\chi$.  The vectors $N^a$ and $N^{a'}$ are in the $\chi$
direction, and we have
\begin{eqnarray}
N^a & = & H\cos\tau\,\left(\frac{\partial}{\partial \chi}\right)^a\,, \\
N^{a'} & = & -H\cos\tau'\,\left(\frac{\partial}{\partial \chi}\right)^{a'}\,.
\end{eqnarray}
By contracting $\hat{G}_{aba'b'}$ in
(\ref{Gab}) with $N^a$, $N^b$, $N^{a'}$ and $N^{b'}$ we
obtain
\begin{equation}
H^4 \cos^2\tau \cos^2\tau'\hat{G}_{\chi\chi\chi\chi}(x,x')
= \frac{4}{9}g^{(1)}+\frac{4}{3}g^{(2)} \equiv \frac{2}{\pi^2\sin^2\chi}
f^{(1)}\,. \label{f1}
\end{equation}
Next, by contracting $\hat{G}_{aba'b'}$ with $N^a$ and $N^{a'}$ and
letting $b=i$ and $b' =i'$, where $i$ and $i'$ are either $\theta$ or 
$\varphi$, we find
\begin{equation}
H^2\cos\tau\cos\tau'\hat{G}_{\chi i\chi i'}(x,x') = [g^{(2)} - g^{(3)}]g_{ii'}\,.
\end{equation}
Since the parallel propagator $P_{ab'}$ takes vectors parallel to
$\partial/\partial\theta$ or $\partial/\partial\varphi$ to vectors parallel
to the same vectors, we have
\begin{equation}
g_{ii'} = \frac{\sin\chi\sin\chi '}{H^2\cos\tau\cos\tau'}
\tilde{\eta}_{ii'}\,.
\end{equation}
Thus, we find 
\begin{equation}
\lim_{\chi'\rightarrow 0}\frac{\cos^{2}\tau\cos^{2}\tau'} 
{\sin\chi\sin\chi '}H^4\hat{G}_{\chi
i\chi i'}(x,x')
= \tilde{\eta}_{ii'}[g^{(2)}-g^{(3)}] \equiv \frac{1}{2\pi^2\sin^2\chi}
\tilde{\eta}_{ii'}
f^{(2)}\,.  \label{f2}
\end{equation}
Finally, we consider the traceless part of $\hat{G}_{iji'j'}$ 
(with respect to $\tilde{\eta}_{ij}$ and $\tilde{\eta}_{i'j'}$)
with $i$, $j$, $i'$ and $j'$
being either $\theta$ or $\varphi$.
Thus, we define
\begin{equation}
\hat{G}^{\rm tl}_{iji'j'}
\equiv \hat{G}_{iji'j'} - \frac{1}{2}\tilde{\eta}_{ij}
\tilde{\eta}^{kl}\hat{G}_{kli'j'} 
- \frac{1}{2}\tilde{\eta}_{i'j'}\tilde{\eta}^{k'l'}
\hat{G}_{ijk'l'} + \frac{1}{4}\tilde{\eta}_{ij}\tilde{\eta}_{i'j'}
\tilde{\eta}^{kl}\tilde{\eta}^{k'l'}
\hat{G}_{klk'l'}\,.
\end{equation}
Then from (\ref{Gab}) we obtain
\begin{eqnarray}
\lim_{\chi'\to 0}\frac{\cos^{2}\tau\cos^{2}\tau '}{\sin^{2}\chi\sin^{2}\chi'}
H^4\hat{G}^{\rm tl}_{iji'j'}(x,x')
& = & g^{(2)}
(\tilde{\eta}_{ii'}
\tilde{\eta}_{jj'}+\tilde{\eta}_{ij'}
\tilde{\eta}_{i'j}-\tilde{\eta}_{ij}
\tilde{\eta}_{i'j'})\nonumber \\
&  \equiv & \frac{1}{4\pi^2\sin^2\chi}f^{(3)}
(\tilde{\eta}_{ii'}
\tilde{\eta}_{jj'}+\tilde{\eta}_{ij'}
\tilde{\eta}_{i'j}-\tilde{\eta}_{ij}
\tilde{\eta}_{i'j'})\,.
\end{eqnarray}
The functions $g^{(1)}$, $g^{(2)}$ and $g^{(3)}$ can be found from
$f^{(1)}$, $f^{(2)}$ and $f^{(3)}$ as
\begin{eqnarray}
g^{(1)} & = & \frac{1}{4\pi^{2}\sin^{2}\chi}[18f^{(1)} - 3f^{(3)}]\,,
\label{g1} \\
g^{(2)} & = & \frac{1}{4\pi^{2}\sin^{2}\chi}f^{(3)}\,, \label{g2}\\
g^{(3)} & = & \frac{1}{4\pi^{2}\sin^{2}\chi} [f^{(3)} - 2f^{(2)}]
\,. \label{g3}
\end{eqnarray}

We begin by calculating the function $f^{(1)}$.  The functions $f^{(2)}$
and $f^{(3)}$ can be obtained by differentiating $f^{(1)}$ with respect
to $\chi$ [see (\ref{nonsing2}) and (\ref{nonsing3})].
By substituting
(\ref{Tchi}) in (\ref{twoP}) and recalling the definition (\ref{f1}) of
$f^{(1)}$ we have
\begin{eqnarray}
f^{(1)}  & = &  
\lim_{\chi'\to 0}\frac{\pi^2\cos\tau\cos\tau'}{2\sin^2\chi'}
D(\tau)D(\tau')\nonumber \\
&& \times \sum_{L=2}^{\infty}\sum_{l=2}^{L}\sum_{m=-l}^{l}
\frac{(c_{L}^{l})^{2}e^{-i(L+1)(\tau-\tau')}}{L(L+1)(L+2)}
\bar{P}_{L}^{l}(\chi )\bar{P}_{L}^{l}(\chi')
|Y_{lm}(\theta ,\varphi )|^2\,,  \label{f11}
\end{eqnarray}
where the function $\bar{P}_{L}^l(\chi)$ is given by (\ref{barP}) and
the constant $c_L^l$ is given by (\ref{CLL}).  Since the associated
Legendre functions ${\rm P}_{\nu}^{-\mu}(\cos\chi')$ behave like
$(\chi')^\mu$ for small $\chi'$, we have
$\bar{P}_L^l(\chi') \sim (\chi')^l$. [Here and below, 
``$f(\chi')\sim g(\chi')$"
means that $\lim_{\chi'\to 0}f(\chi')/g(\chi') =\,{\rm const}$.] 
Then, we find that only the $l=2$
terms contribute in the infinite sum (\ref{f11}).  
The explicit expression
for $\bar{P}_L^2(\chi)$ can be obtained as follows.  By letting $l=2$ in
(\ref{barP}) we have
\begin{equation}
\bar{P}_L^2(\chi) = \left[ \frac{(L+1)(L+3)!}{(L-2)!}\right]^{\frac{1}{2}}
(\sin\chi)^{-1/2}{\rm P}_{L+1/2}^{-5/2}(\cos\chi)\,.
\end{equation} 
The function ${\rm P}_{L+1/2}^{-5/2}(\cos\chi)$ can be found by noting
that~\cite{Grad}
\begin{equation}
{\rm P}_{L+1/2}^{-1/2}(\cos\chi ) =
\left(\frac{2}{\pi\sin\chi}\right)^{\frac{1}{2}}\frac{\sin
(L+1)\chi}{L+1}
\end{equation}
and using the lowering operator for the associated Legendre functions:
\begin{equation}
{\rm P}_\nu^{-\mu-1}(x) = 
\frac{1}{(\nu-\mu)(\nu+\mu+1)}\left[
\sqrt{1-x^2}\frac{d\ }{dx}
+ \frac{\mu x}{\sqrt{1-x^2}}\right]{\rm P}_{\nu}^{-\mu}(x)\,.
\end{equation}
The result is
\begin{equation}
\bar{P}_L^2(\chi) 
= \left[\frac{2}{\pi(L-1)L(L+2)(L+3)}\right]^{1/2}
\left(\frac{d}{d\chi}-\cot\chi\right )\frac{d}{d\chi}\left[\frac{\sin
(L+1)\chi}{\sin\chi}\right]\,.
\end{equation}
This can be used to show that
\begin{equation}
\lim_{\chi'\to 0} \frac{1}{\sin^2\chi'}
\bar{P}_L^2(\chi')\bar{P}_L^2(\chi)
= \frac{2(L+1)}{15\pi}
\left(\frac{d}{d\chi}-\cot\chi\right )\frac{d}{d\chi}\left[
\frac{\sin (L+1)\chi}{\sin\chi}\right]\,.  \label{lim}
\end{equation}
We also note that
\begin{equation}
\sum_{m=-2}^2 Y_{2m}(\theta,\varphi)\overline{Y_{2m}(\theta',\varphi')}
= \frac{5}{4\pi}{\rm P}_2(\cos\gamma)
= \frac{5}{8\pi}(3\cos^2\gamma-1)\,,  \label{presum}
\end{equation}
where $\gamma$ is the angle between the points on the unit $S^2$ with
coordinates $(\theta,\varphi)$ and $(\theta',\varphi')$:
\begin{equation}
\cos\gamma = \cos\theta \cos\theta' + \sin\theta\sin\theta'
\cos(\varphi-\varphi')\,. \label{pres1}
\end{equation}
Hence,
\begin{equation}
\sum_{m=-2}^2 |Y_{2m}(\theta,\varphi)|^2 = \frac{5}{4\pi}\,. \label{summa}
\end{equation}
Substituting (\ref{lim}), (\ref{summa}) and using $c_L^2$ found from
(\ref{CLL}),
\begin{equation}
(c_L^2)^2 = \frac{12}{L(L+1)^2(L+2)}\,,
\end{equation}
in (\ref{f11}), we obtain
\begin{eqnarray}
f^{(1)}  & = & \cos\tau\cos\tau' D(\tau)D(\tau')\nonumber \\
&& \times \sum_{L=2}^{\infty}
\frac{e^{-i(L+1)(\tau-\tau')}}{L^2(L+1)^2(L+2)^2}
\left(\frac{d}{d\chi}-\cot\chi\right )\frac{d}{d\chi}\left[
\frac{\sin (L+1)\chi}{\sin\chi}\right]\,. \label{nonsing1}
\end{eqnarray}
This can be written as
\begin{eqnarray}
f^{(1)} & = & \cos\tau\cos\tau'D(\tau )D(\tau ')\nonumber \\
&& \times \sum_{L=2}^{L=\infty}\frac{1}{L^{2}(L+1)^{2}(L+2)^{2}}
\times\left\{\left [ \frac{3}{2i\sin^{3}\chi}  -
\frac{(L+1)^{2}+2}{2i\sin\chi}\right ] (X^{L+1} -\tilde{X}^{L+1} )
\right. \nonumber \\
&& \left. \ \ \ \ \ \ -\frac{3(L+1)\cos\chi}{2\sin^{2}\chi}
(X^{L+1} +\tilde{X}^{L+1})\right\}\,, \label{series}
\end{eqnarray}
where
\begin{eqnarray}
X & = & e^{-i(\tau -\tau '-i\epsilon)+i\chi}\,, \\
\tilde{X} & = & e^{-i(\tau -\tau'-i\epsilon)-i\chi}\,.
\end{eqnarray}
Note that we have inserted the convergence factor $e^{-\epsilon}$ with
$\epsilon > 0$ and that $|X| = |\tilde{X}| = e^{-\epsilon}$.  Since 
the radius of convergence is one for the power series in (\ref{series}),
this equation can be differentiated with respect to $\chi$, $\tau$ or $\tau'$,
and the limit $\chi \to 0$, which
we will need later, can be taken under the 
summation sign once this convergence factor is in place,
because the series (before and after these operations)
will be absolutely convergent.

The series in (\ref{series}) can be evaluated by manipulating the
geometric series $\sum_{L=0}^{\infty}x^L = (1-x)^{-1}$. 
The result is
\begin{eqnarray}
f^{(1)} & = & \cos\tau\cos\tau' 
D(\tau )D(\tau ')\nonumber \\
&& \times \left\{\frac{A(X)-A(\tilde{X})}{2i\sin^{3}\chi} -
\frac{\left[B(X) + B(\tilde{X})\right]\cos\chi}{2\sin^{2}\chi} 
- \frac{C(X)-C(\tilde{X})}{2i\sin\chi}\right\}\,,
\end{eqnarray}
where
\begin{eqnarray}
A(x) & = & \frac{9}{4}\left(x - \frac{1}{x}\right)\log (1-x) - \frac{69}{16}x-
\frac{x^2}{12} \nonumber \\
&& - \left[\frac{3}{4}\left(x+\frac{1}{x} \right)+3\right]\int_0^x
\log (1 - \xi)\frac{d\xi}{\xi}\,, \\
B(x) & = & 
\left\{ \frac{3}{2}\left(x+\frac{1}{x}\right) - 3\right\}\log (1-x)
+ \frac{9}{4} - \frac{33}{16}x - \frac{x^2}{6}\nonumber \\ 
&& - \frac{3}{4}\left(x-\frac{1}{x}\right)\int_0^x\log(1-\xi)\frac{d\xi}{\xi}
\,,\\
C(x) & = & \frac{7}{4}\left(x-\frac{1}{x}\right)\log (1-x) -
\frac{49}{16}x - \frac{x^2}{6}\nonumber \\
&&  -\left\{ \frac{3}{4}\left(x+\frac{1}{x}\right) + 2\right\}
\int_0^x\log (1-\xi)\frac{d\xi}{\xi}\,.
\end{eqnarray}
Application of the differential operator $D(\tau)D(\tau')$ is made slightly
easier by using 
\begin{eqnarray}
&& D(\tau )D(\tau ')\left[\left(X\pm\frac{1}{X}\right)p(X )\right]\nonumber \\
&& =  p(X )D(\tau )D(\tau ')\left(X\pm\frac{1}{X}\right)
+\left[\frac{\partial p(X )}{\partial \tau}\right]D(\tau')
\left(X\pm\frac{1}{X}\right)\nonumber \\
&& + \left[\frac{\partial p(X )}{\partial \tau'}\right]D(\tau)
\left(X\pm\frac{1}{X}\right) 
+ \left(X\pm\frac{1}{X}\right)\frac{\partial^{2}p(X )}
{\partial\tau \partial\tau '}\,,
\end{eqnarray}
where $p(X)$ is an arbitrary function, and a similar formula with
$X$ replaced by $\tilde{X}$, because one can then use the identities
$\cos \tau D(\tau)e^{-i\tau}= -i$ and 
$\cos \tau' D(\tau')e^{i\tau'}= i$.  We find
\begin{eqnarray}
f^{(1)} & = & 
\left[-\frac{3\cos\chi}{4\sin^{3}\chi} +
\left(\frac{1}{\sin\chi} - \frac{3}{2\sin^{3}\chi}\right)\sin\tau\sin\tau'
\right]\int_{-\chi}^{\chi}\log \left[ 
1 - e^{-i(\tau -\tau'-i\epsilon)+ i\xi }\right] 
d\xi\nonumber \\
&& + \left(-\frac{1}{4} +
\frac{3}{4\sin^{2}\chi} + \frac{3\cos\chi\sin\tau\sin\tau'}
{2\sin^{2}\chi}\right)\log [(1-X)(1-\tilde{X})]\nonumber \\
&& +\left[\frac{3\cos\chi}{4\sin^{3}\chi}\cos 
(\tau + \tau') 
+ \frac{1}{\sin\chi} - \frac{3}{2\sin^{3}\chi}
\right]\sin(\tau-\tau')\log\frac{1-X}{1-\tilde{X}}\nonumber \\
&& + \frac{3}{4} - \frac{3}{2\sin^{2}\chi} +
\left(\frac{3}{8\sin^{2}\chi} - \frac{1}{8}\right)(e^{-2i\tau} 
+ e^{2i\tau'}) \nonumber \\
&& +  
\frac{\cos\chi}{\sin^{2}\chi}\left[\frac{3}{4}\cos (\tau + \tau ') -
\frac{3}{2}e^{-i(\tau -\tau ')}\right]\,. \label{f1res}
\end{eqnarray}

To find $f^{(2)}$ defined by (\ref{f2}) we first note that
\begin{equation}
f^{(2)}\tilde{\eta}_{ii'}
= 2\pi^2\cos^2\tau\cos^2\tau' \sin\chi \lim_{\chi '\rightarrow 0}
\frac{1}{\sin\chi'} 
H^4\hat{G}_{\chi i\chi i'}(x,x')\,.  \label{f22} 
\end{equation}
The component $T^{({\rm v};Llm)}_{\chi i}$ behaves like
$\chi^l$ for small $l$ whereas $T^{({\rm s};Llm)}_{\chi i} \sim \chi^{l-1}$.
{}From (\ref{twoP}) and (\ref{f22}) we find that the tensors
$T^{({\rm v};Llm)}_{ab}$ do not contribute to $f^{(2)}$ because
\begin{equation}
\lim_{\chi'\to 0}\frac{1}{\sin\chi'}T^{({\rm v};Llm)}_{\chi i'}(\chi',\theta',
\varphi') = 0
\end{equation}
for all $l \geq 2$.  As in the case for $f^{(1)}$ we find that only
the $l=2$ modes of $T^{({\rm s};Llm)}_{ab}$ contribute.  Thus, we have
\begin{eqnarray}
\tilde{\eta}_{ii'}f^{(2)}  & = &  
\frac{\pi^2}{18} \cos\tau \cos\tau' D(\tau)D(\tau')
\sin\chi \left( \frac{\partial\ }{\partial \chi} + \cot\chi\right)
\lim_{\chi'\to 0}
\frac{1}{\sin\chi'}\left(\frac{\partial\ }{\partial\chi'} + \cot\chi'\right)
\nonumber \\
&& \times \sum_{L=2}^{\infty}
\frac{(c_{L}^{2})^{2}e^{-i(L+1)(\tau-\tau')}}{L(L+1)(L+2)}
\bar{P}_{L}^{2}(\chi )\bar{P}_{L}^{2}(\chi')
\lim_{\gamma\to 0}\tilde{D}_i\tilde{D}_{i'}
\frac{5}{8\pi}(3\cos^2\gamma-1)\,.
\label{f222}
\end{eqnarray}
Since $\bar{P}_L^2(\chi') \sim (\chi')^2$, it can readily be seen that
\begin{equation}
\lim_{\chi'\to 0} \frac{1}{\sin\chi'}
\left(\frac{\partial\ }{\partial\chi'} + \cot\chi'\right)
\bar{P}_L^2(\chi')\bar{P}_L^2(\chi)
= 3 \lim_{\chi'\to 0}\frac{1}{\sin^2\chi'}
\bar{P}_L^2(\chi')\bar{P}_L^2(\chi)\,.  \label{pp}
\end{equation}
We also find by using (\ref{pres1}) that
\begin{equation}
\lim_{\gamma\to 0}\tilde{D}_i\tilde{D}_{i'}
\frac{5}{8\pi}(3\cos^2\gamma-1)
= \frac{15}{4\pi}\tilde{\eta}_{ii'}  \label{DDi}
\end{equation}
in the limit $(\theta,\varphi)\to (\theta',\varphi')$.  By comparing (\ref{DDi})
and (\ref{summa}) and using (\ref{pp}), and also comparing (\ref{f11}) and
(\ref{f222}), we find
\begin{equation}
f^{(2)} = \frac{\partial\ }{\partial\chi}\left[ \sin\chi f^{(1)}\right]\,.
\label{nonsing2}
\end{equation} 
By a tedious but straightforward calculation we obtain
\begin{eqnarray}
f^{(2)} & = & \left( \frac{3}{2\sin^{3}\chi} - \frac{3}{4\sin\chi} +
\frac{3\cos\chi}{\sin^{3}{\chi}}\sin\tau\sin\tau
'\right)\int_{-\chi}^{\chi}\log \left[
1 - e^{-i(\tau - \tau'-i\epsilon) + i\xi}\right]d\xi \nonumber \\
&& +\left[-\frac{\cos\chi}{4}-
\frac{3\cos\chi}{2\sin^{2}\chi} + \left(1 -
\frac{3}{\sin^{2}\chi}\right)\sin\tau\sin\tau '\right]\log 
\left[(1-X)(1-\tilde{X})\right]\nonumber \\
&& + \left[\left(\frac{3}{4\sin\chi}-\frac{3}{2\sin^{3}\chi}\right)\cos 
(\tau + \tau')
+ \frac{3\cos\chi}{\sin^{3}\chi}
\right]\sin(\tau-\tau')\log\frac{1-X}{1-\tilde{X}}\nonumber \\
&& + \frac{\cos\chi}{2} +
\frac{3\cos\chi}{\sin^{2}\chi} - \left[\frac{3\cos\chi}{4\sin^{2}\chi} +
\frac{\cos\chi}{8}\right](e^{-2i\tau}+e^{2i\tau'})\nonumber \\
&& +\left[\frac{3}{4} - \frac{3}{2\sin^{2}\chi}\right]\cos (\tau + \tau') 
+ \left(\frac{3}{\sin^{2}\chi} - 1\right)e^{-i(\tau -\tau')} \,.
\label{f2res}
\end{eqnarray}

We need to evaluate $\hat{G}^{\rm tl}_{iji'j'}$ in order to find $f^{(3)}$. As
in the case for $f^{(2)}$, the modes $T^{({\rm v};Llm)}_{ab}$ do not contribute
because
\begin{equation}
\lim_{\chi'\to 0}\frac{1}{\sin^2\chi'}T^{({\rm v};Llm)}_{i'j'}(\chi',\theta',
\varphi') = 0
\end{equation}
for all $l \geq 2$, and only 
the $l=2$ modes of $T^{({\rm s};Llm)}_{ab}$ 
contribute.  By defining the traceless
part of $T^{({\rm s};Llm)}_{ij}$ given by (\ref{misprint}) as
\begin{equation}
T_{ij}^{({\rm tl};Llm)} = 
c_{L}^{l} F_L^l(\chi )\left[\tilde{D}_{i}\tilde{D}_{j} +
\frac{l(l+1)}{2}\tilde{\eta}_{ij}\right]Y_{lm}(\theta ,\varphi )
\end{equation}
we have 
\begin{eqnarray}
H^4\hat{G}^{\rm tl}_{iji'j'}(x,x')
& = &   
\sec\tau\sec\tau'D(\tau)D(\tau')
\lim_{\chi'\to 0}\sum_{L=2}^\infty 
\sum_{m=-2}^2 \nonumber \\
&& \times \frac{e^{-i(L+1)(\tau-\tau')}}{L(L+1)(L+2)}
T_{ij}^{({\rm tl};L2m)}(\chi,\theta,\varphi)
\overline{T_{i'j'}^{({\rm tl};L2m)}(\chi',\theta,\varphi)}\,.
\end{eqnarray}
Hence,
\begin{eqnarray}
K_{iji'j'}f^{(3)} & = & 4\pi^2 \cos\tau\cos\tau' D(\tau)D(\tau') \nonumber \\
&& \times \lim_{\chi'\to 0}\sum_{L=2}^\infty (c_L^2)^2
\frac{e^{-i(L+1)(\tau-\tau'-i\epsilon)}}
{L(L+1)(L+2)}\frac{F_L^2(\chi)F_L^2(\chi')}{\sin^2\chi'}\nonumber \\
&& \times \lim_{\gamma\to 0}
\left(\tilde{D}_i\tilde{D}_j + 3\tilde{\eta}_{ij}\right)
\left(\tilde{D}_{i'}\tilde{D}_{j'} + 3\tilde{\eta}_{i'j'}\right)
\frac{5}{8\pi}(3\cos^2\gamma-1)\,,
\label{f3}
\end{eqnarray}
where 
\begin{equation}
K_{iji'j'} \equiv \tilde{\eta}_{ii'}\tilde{\eta}_{jj'}
+ \tilde{\eta}_{ij'}\tilde{\eta}_{ji'} 
- \tilde{\eta}_{ij}\tilde{\eta}_{i'j'}\,.
\end{equation}
Again, since $\bar{P}_L^2(\chi')\sim (\chi')^2$ for small $\chi'$, we find
\begin{equation}
\lim_{\chi'\to 0} F_L^2(\chi)F_L^2(\chi')
= \frac{1}{24}
\left\{ \frac{\partial\ }{\partial\chi}\left[
\sin^2\chi\left( \frac{\partial\ }{\partial \chi} + \cot\chi\right)\right]
- 3\right\}
\lim_{\chi'\to 0}\frac{\bar{P}_L^2(\chi')}{\sin^2\chi'}
\bar{P}_L^2(\chi)\,.  \label{OK}
\end{equation} 
We also find 
that in the limit $\gamma \to 0$
\begin{eqnarray}
\tilde{D}_i\tilde{D}_j \frac{5}{8\pi}(3\cos^2\gamma -1)
& \to &  - \frac{15}{4\pi}\tilde{\eta}_{ij} 
\,,\\
\tilde{D}_{i'}\tilde{D}_{j'}\frac{5}{8\pi}(3\cos^2\gamma-1)
& \to &  - \frac{15}{4\pi} \tilde{\eta}_{i'j'} 
\,,\\
\tilde{D}_i\tilde{D}_j\tilde{D}_{i'}\tilde{D}_{j'} 
\frac{5}{8\pi}(3\cos^2\gamma-1)
& \to & \frac{15}{4\pi}\left[ \tilde{\eta}_{ii'} 
\tilde{\eta}_{jj'} + \tilde{\eta}_{ij'}\tilde{\eta}_{ji'}
+ 2\tilde{\eta}_{ij}\tilde{\eta}_{i'j'}\right]\,.
\end{eqnarray}
These together with (\ref{summa}) imply that
\begin{equation}
\lim_{\gamma\to 0}
\left(\tilde{D}_i\tilde{D}_j + 3\tilde{\eta}_{ij}\right)\left(
\tilde{D}_{i'}\tilde{D}_{j'} + 3\tilde{\eta}_{i'j'}\right)
\frac{5}{8\pi}(3\cos^2\gamma-1)
= \frac{15}{4\pi}K_{iji'j'}\,. \label{good}
\end{equation}
By substituting (\ref{OK}) and (\ref{good}) in (\ref{f3}) we find
\begin{equation}
f^{(3)} = \frac{\partial\ }{\partial\chi}\left[ \sin\chi f^{(2)}\right]
- 3f^{(1)}\,. \label{nonsing3}
\end{equation} 
Then, a tedious but straightforward calculation gives the following result:
\begin{eqnarray}
f^{(3)} & = & 
\frac{\sin^{2}\chi\cos\tau\cos\tau '}{2[\cos (\tau -\tau'-i\epsilon)-
\cos\chi ]} \nonumber \\
&& -\left(\frac{3\cos\chi}{4\sin^{3}\chi} +
\frac{3}{2\sin^{3}\chi}\sin\tau\sin\tau '\right)\int_{-\chi}^{\chi} \log 
\left[1 - e^{-i(\tau-\tau'-i\epsilon) + i\xi}\right] d\xi\nonumber \\
&& +\left[-\frac{1}{4} + \frac{3}{4\sin^{2}\chi} + \frac{\sin^{2}\chi}{2} +
\left(\cos\chi +
\frac{3\cos\chi}{2\sin^{2}\chi}\right)\sin\tau\sin\tau '\right]\nonumber \\
&& \ \ \  \times \log
[(1-X)(1-\tilde{X})]\nonumber \\
&&
+ \left[\frac{3\cos\chi}{4\sin^{3}\chi}\cos (\tau + \tau ')
-\frac{3}{2\sin^{3}\chi}\right] 
\sin(\tau-\tau') \log\frac{1-X}{1-\tilde{X}}\nonumber \\
&& -\frac{1}{4} -
\frac{3}{2\sin^{2}\chi} -\frac{3\sin^{2}\chi}{4}
+ \frac{3\cos\chi}{4\sin^{2}\chi}\cos (\tau + \tau ') \nonumber \\
&& + \left(\frac{\sin^{2}\chi}{4} -\frac{1}{8} +
\frac{3}{8\sin^{2}\chi}\right)(e^{-2i\tau} + e^{2i\tau '}) \nonumber \\
&& - \left(\frac{3\cos\chi}{2\sin^{2}\chi} +
\cos\chi\right)e^{-i(\tau -\tau ')}\,. \label{f3res}
\end{eqnarray}
We can readily 
find the functions $g^{(1)}$, $g^{(2)}$ and $g^{(3)}$ which appear
in the expression (\ref{Gab}) for the two-point function by substituting 
(\ref{f1res}), (\ref{f2res}) and (\ref{f3res}) in (\ref{g1}), (\ref{g2})
and (\ref{g3}). 

The flat-space limit of the two-point function (\ref{Gab}) can be obtained
by defining
$\chi = H r$, $\tau = Ht$ and $\tau'=Ht'$ and letting $H\to 0$.  The two
spacetime points are at $t$ and $t'$, respectively, and their spatial
distance is $r$.  The tensors $N^a$, $N^{a'}$, $g_{ab'}$, $g_{ab}$ and
$g_{a'b'}$ tend to the corresponding tensors in flat spacetime.
By writing the flat-space limit of 
$H^2 g^{(i)}$ as $g^{(i)}_{\rm flat}$ we find
\begin{eqnarray}
g^{(1)}_{\rm flat} & = &  \frac{1}{4\pi^2}
\left(-\frac{3}{r^{2}-T^{2}} + \frac{11}{r^{2}} -
\frac{15T^{2}}{2r^{4}} - \frac{27T}{4r^{3}}\log\frac{T+r}{T-r} +
\frac{15T^{3}}{4r^{5}}\log\frac{T+r}{T-r}\right), \label{flat1}\\
g^{(2)}_{\rm flat} & = &  \frac{1}{4\pi^2}\left(
\frac{1}{r^{2}-T^{2}} - \frac{5}{3r^{2}} -
\frac{T^{2}}{2r^{4}} + \frac{3T}{4r^{3}}\log\frac{T+r}{T-r} +
\frac{T^{3}}{4r^{5}}\log\frac{T+r}{T-r}\right), \label{flat2} \\
g^{(3)}_{\rm flat} & = & \frac{1}{4\pi^2}\left(
\frac{1}{r^{2}-T^{2}} - \frac{7}{3r^{2}} -
\frac{5T^{2}}{2r^{4}} + \frac{3T}{4r^{3}}\log\frac{T+r}{T-r} +
\frac{5T^{3}}{4r^{5}}\log\frac{T+r}{T-r}\right),  \label{flat3}
\end{eqnarray}
where we have defined $T \equiv t'-t + i\epsilon$.  These limits agree with
the results obtained directly in Minkowski spacetime.  The detail of the 
latter calculation is given in Appendix A.

\section{Some properties of the two-point function}

\setcounter{equation}{0}

The expressions (\ref{f1res}), (\ref{f2res}) and (\ref{f3res}) for
the functions $f^{(1)}$, $f^{(2)}$ and $f^{(3)}$, respectively,
may appear singular at
$\sin\chi = 0$.  However, in fact they behave like $\sin^2\chi$ as is
clear from (\ref{nonsing1}), (\ref{nonsing2}) and (\ref{nonsing3}).
Let us define
\begin{equation}
F_0 \equiv \lim_{\chi\to 0} \frac{f^{(1)}}{\sin^2\chi}\,.
\end{equation}
We find\footnote{We obtained this limit by first letting
$\chi \to 0$ in (\ref{nonsing1}) divided by $\sin^2\chi$
and then performing the $L$ summation.
We then verified that this result agrees with the $\chi\to 0$
limit of (\ref{f1res}) divided by $\sin^2\chi$ by Maple.}
\begin{eqnarray}
F_0 & = & - \frac{1}{12} + \frac{2}{15}e^{-i(\tau-\tau')} + 
\frac{1}{20}(e^{-2i\tau} + e^{2i\tau'}) \nonumber \\
&& - \frac{\cos\tau\cos\tau'}{30[1-\cos(\tau-\tau'-i\epsilon)]}
+ \left( \frac{1}{5} - \frac{4}{15}\sin\tau\sin\tau'\right)
\log\left[ 1-e^{-i(\tau-\tau'-i\epsilon)}\right]\,. \label{F0}
\end{eqnarray}
Then, by using (\ref{nonsing2}) and (\ref{nonsing3}) we obtain
\begin{eqnarray}
\lim_{\chi\to 0}\frac{f^{(2)}}{\sin^2\chi} & = & 3F_0\,,\\
\lim_{\chi\to 0}\frac{f^{(3)}}{\sin^2\chi} & = & 6F_0\,.
\end{eqnarray}
Thus, we find the $\chi\to 0$ limit of the two-point function as 
\begin{equation}
\lim_{\chi\to 0}\hat{G}_{aba'b'} = 
\frac{3F_0}{2\pi^2}\left( g_{aa'}g_{bb'} + g_{ab'}g_{a'b} -
\frac{2}{3}g_{ab}g_{a'b'}\right)\,.
\end{equation}
Indeed there is no singularity in the limit
$\chi\to 0$ as long as $\tau\neq \tau'$.
If $|\tau-\tau'| \ll 1$ we have
\begin{equation}
\lim_{\chi\to 0}\hat{G}_{aba'b'}
\approx - \frac{1}
{10\pi^2}\frac{\cos^2\tau}{(\tau-\tau'-i\epsilon)^2}
\left( g_{aa'}g_{bb'} + g_{ab'}g_{a'b}
-\frac{2}{3}g_{ab}g_{a'b'}\right)\,.  \label{OK2}
\end{equation}
This must be reproduced by the flat-space result
since it concerns with
the short-distance behaviour of the two-point function.  We indeed find from
(\ref{flat1}), (\ref{flat2}) and (\ref{flat3}) that
$g^{(1)}_{\rm flat} \to 0$, $g^{(2)}_{\rm flat} \to -(10\pi^2 T^2)^{-1}$
and $g^{(3)}_{\rm flat} \to 0$ as $r\to 0$, thus reproducing
(\ref{OK2}) with $T = (H\cos\tau)^{-1}(\tau' - \tau + i\epsilon)$. 

Next we consider the $\chi\to \pi$ limit.  It can easily be obtained from the
$\chi\to 0$ limit as follows.   Define $\psi \equiv \pi - \chi$.  
The variable $\chi$ can be replaced by $\psi$ if
a factor of $(-1)^L$ is inserted in each term in (\ref{nonsing1}).  Hence, the 
$\chi\to \pi$ limit is found from the $\chi \to 0$ limit by
letting $\tau \to \tau + \pi$ and multiplying by $-1$.  
The $\chi\to \pi$ limit 
of $f^{(1)}/\sin^2\chi$ thus obtained is 
\begin{equation}
\lim_{\chi\to \pi}\frac{f^{(1)}}{\sin^2\chi}
= F_1\,,
\end{equation}
where
\begin{eqnarray}
F_1 & \equiv & 
\frac{1}{12} + \frac{2}{15}e^{-i(\tau-\tau')} -
\frac{1}{20}(e^{-2i\tau} + e^{2i\tau'}) \nonumber \\
&& - \frac{\cos\tau\cos\tau'}{30[1+\cos(\tau-\tau'-i\epsilon)]}
- \left( \frac{1}{5} + \frac{4}{15}\sin\tau\sin\tau'\right)
\log\left[ 1+e^{-i(\tau-\tau'-i\epsilon)}\right]\,. \label{F1}
\end{eqnarray} 
In a manner similar to the $\chi \to 0$ case, we have
\begin{eqnarray}
\lim_{\chi\to 0}\frac{f^{(2)}}{\sin^2\chi} & = & -3F_1\,,\\
\lim_{\chi\to 0}\frac{f^{(3)}}{\sin^2\chi} & = & 6F_1\,.
\end{eqnarray}
Then we find
\begin{eqnarray}
\lim_{\chi\to \pi}\hat{G}_{aba'b'} 
& = & 
\frac{3F_1}{2\pi^2}\left[
(g_{aa'} + 2N_{a}N_{a'})(g_{bb'}+2N_{b}N_{b'}) +
(g_{ab'} + 2N_{a}N_{b'})(g_{ba'}+2N_{b}N_{a'})\right]\nonumber \\
&& \ \ \ - \frac{F_1}{\pi^2}g_{ab}g_{a'b'}\,. \label{welld}
\end{eqnarray}
Now, recall that the geodesic connecting the two points at $\chi=0$ and $\pi$
on $S^3$ is not uniquely determined. Hence the tensors 
$N^a$, $N^{a'}$ and $g_{aa'}$ are not well defined unlike the metrics
$g_{ab}$ and $g_{a'b'}$.  However, the combination
$g_{aa'} + 2N_a N_{a'}$ is well defined because the vector
$({g^a}_{a'}+ 2N^a N_{a'})V^{a'}$ for a given vector $V^{a'}$ is 
independent of
the geodesic chosen to define it.  Equation (\ref{welld}) shows that 
the $\chi\to \pi$ limit of
the two-point function is well defined.

As we mentioned in the introduction, the physical two-point function 
in the spatially flat coordinate system with
the metric
\begin{equation}
ds_{\rm flat}^2 = - dt^2 + e^{2Ht}\left( dx^2 + dy^2 + dz^2\right)
\label{flat}
\end{equation}
exhibits logarithmic growth as the coordinate distance of the two points
$(t,x,y,z)$ and $(t',0,0,0)$ defined by
$(x^2+y^2+z^2)^{1/2}$ becomes large.  The two-point function in the
covariant gauge has a similar behaviour. Although this growth has been shown 
to be 
a gauge artefact, it is interesting to have a two-point function which does
not exhibit this behaviour.

The two-point function obtained in this paper is indeed bounded
[except near the 
light-cone singularity with $\chi = \pm(\tau-\tau')$] as long as $|\tau|$ or
$|\tau'|$ is fixed at a value less than $\pi/2$. 
As we show in Appendix B, the limit $(x^2+y^2+z^2)^{1/2}\to \infty$ with
$t$ fixed for the spatially flat coordinate system 
corresponds to the limit 
$\chi\to \pi$ and $\tau \to \pi/2$ with $\tau'$ held fixed.
This limit can readily be found
from (\ref{F1}) as
\begin{equation} 
F_1 \to  
\frac{2}{15} - \frac{2i}{15}e^{i\tau'} -
\frac{1}{20}e^{2i\tau'}
- \left( \frac{1}{5} + \frac{4}{15}\sin\tau'\right)
\log\left( 1-ie^{i\tau'}\right)\,.
\end{equation} 
This clearly shows that the logarithmic
growth of the two-point functions in various other gauges in this limit is a
gauge artefact.
One can send both points to infinity 
while keeping them on the same spatial
section in the spatially flat coordinate system by letting
$\tau = \tau' \to \pi/2$ and $\chi \to 0$ 
with $\cos \tau/\sin^2\chi$ kept finite.  In this limit we find
\begin{equation}
\frac{f^{(1)}}{\sin^2\chi} \approx -\frac{13}{900} - \frac{1}{15}\log\chi\,.
\end{equation}
Thus, the two-point function diverges logarithmically.  
This fact does not invalidate our assertion that the logarithmic grow is
a gauge artefact because one point can
always be kept at a fixed point using de~Sitter invariance.

It is known~\cite{Hawketal}
that the corresponding two-point 
function in the hyperbolic coordinate system with the metric
\begin{equation}
ds_{\rm hyper}^2 = H^{-2}\left\{ -d\eta^2
+ \sinh^2\eta\left[ d\zeta^2 + \sinh^2\zeta (d\theta^2 + \sin^2\theta
d\varphi^2)\right]\right\}  \label{hyperb}
\end{equation}
does not grow for large coordinate distance, i.e. when
one point is at $\zeta = 0$ and the other point has large  $\zeta$.
However, this coordinate system does not cover the region where the
spatially flat two-point function grows (see Appendix B).  
The limit $\zeta\to\infty$ in fact corresponds to the large $t$ limit with
$(x^2+y^2+z^2)^{1/2} \to H^{-1}$. 

\section*{Appendix A. The flat-space two-point function}

\setcounter{equation}{0}
\renewcommand{\theequation}{A\arabic{equation}}

In this Appendix we calculate the graviton two-point function in
the gauge adopted in this paper in Minkowski spacetime.  The result is
given by the right-hand sides of (\ref{flat1}), (\ref{flat2}) and 
(\ref{flat3}).

We denote the flat-space two-point function 
for the gravitational perturbation
$h_{\mu\nu}$ in Minkowski spacetime in the gauge
$h_{0\mu} = 0$, $\partial_a h^{ab} = 0$ and ${h^{a}}_a = 0$
by $\Delta_{aba'b'}(x,x')$ with
$x\equiv (t,{\bf r})$ and $x'\equiv (t',{\bf r}')$.  
(Here, $a$ and $b$ are spatial indices in the
cartesian coordinate system.)  Then
\begin{eqnarray}
\hat{\Delta}_{aba'b'} & \equiv & 
(16\pi G)^{-1} \Delta_{aba'b'} \nonumber \\
& = & \int\frac{d^3{\bf k}}{(2\pi)^3 2k}
H_{aba'b'}({\bf k}) 
e^{-ik(t-t'-i\epsilon)+i{\bf k}\cdot({\bf r} - {\bf r}')}\,,
\end{eqnarray}
where
\begin{eqnarray}
H_{aba'b'}({\bf k}) & \equiv &
\left( \delta_{aa'} - \frac{k_a k_{a'}}{k^2}\right)
\left( \delta_{bb'} - \frac{k_b k_{b'}}{k^2}\right)
+ \left( \delta_{ab'} - \frac{k_a k_{b'}}{k^2}\right)
\left( \delta_{ba'} - \frac{k_b k_{a'}}{k^2}\right)\nonumber \\
&& - \left( \delta_{ab} - \frac{k_a k_{b}}{k^2}\right)
\left( \delta_{a'b'} - \frac{k_{a'} k_{b'}}{k^2}\right)
\end{eqnarray}
with $k \equiv \|{\bf k}\|$.
Define $T \equiv t' - t + i\epsilon$ and
\begin{eqnarray}
E_n & \equiv & 
\int\frac{d^3{\bf k}}{(2\pi)^3 2k^{n+1}}
e^{ikT+i{\bf k}\cdot({\bf r} - {\bf r}')}\nonumber \\
 & = & \frac{1}{8\pi^2 i r}\int_{\alpha}^\infty \frac{dk}{k^n}
(e^{ik(T+r)} - e^{ik(T-r)})\,,  \label{En}
\end{eqnarray}
where $\alpha$ is an infrared cut-off, which is necessary if
$n \geq 1$.
We convert the factors of $k_a$ to space derivatives with the following 
result:
\begin{eqnarray}
\hat{\Delta}_{aba'b'}
& = & (\delta_{aa'}\delta_{bb'} + \delta_{ab'}\delta_{ba'}
- \delta_{ab}\delta_{a'b'}) E_0 \nonumber \\
&& + (\delta_{aa'}\partial_{b}\partial_{b'}
+ \delta_{bb'}\partial_{a}\partial_{a'}
+ \delta_{ab'}\partial_{b}\partial_{a'}
+ \delta_{ba'}\partial_{b}\partial_{a'}
- \delta_{ab}\partial_{a'}\partial_{b'}
- \delta_{a'b'}\partial_{a}\partial_{b}) E_2 \nonumber \\
&& + \partial_a \partial_b \partial_{a'}\partial_{b'}
E_4\,, \nonumber
\end{eqnarray}
where the derivatives $\partial_a$ and $\partial_{a'}$ are both with respect
to ${\bf r}$ (and not ${\bf r}'$).

Define $\hat{r}_a \equiv (r_a - r'_a)/r$ and 
$\hat{r}_{a'} \equiv (r_{a'} - r'_{a'})/r$
with $r \equiv \|{\bf r}-{\bf r}'\|$.  By using
\begin{equation}
\frac{\partial f}{\partial r_a} = \frac{df}{dr}\hat{r}_a
\end{equation}
for any function $f$ of $r$ and
\begin{equation}
\frac{\partial \hat{r}_a}{\partial r_b} 
= \frac{\delta_{ab} - \hat{r}_a\hat{r}_b}{r}\,,
\end{equation}
we find
\begin{eqnarray}
\Delta_{aba'b'}
& = & C_1 (\delta_{aa'}\delta_{bb'} + \delta_{ab'}\delta_{ba'})
+ C_2 \delta_{ab}\delta_{a'b'}\nonumber \\
&& 
+ C_3 (\delta_{aa'}\hat{r}_b\hat{r}_{b'} + \delta_{ab'}\hat{r}_b\hat{r}_{a'}
+ \delta_{ba'}\hat{r}_a\hat{r}_{b'} + \delta_{bb'}\hat{r}_a\hat{r}_{a'}
- \delta_{ab}\hat{r}_{a'}\hat{r}_{b'}
- \delta_{a'b'}\hat{r}_{a}\hat{r}_{b})\nonumber \\
&& + C_4 (\delta_{aa'}\hat{r}_b\hat{r}_{b'} + \delta_{ab'}\hat{r}_b\hat{r}_{a'}
+ \delta_{ba'}\hat{r}_a\hat{r}_{b'} + \delta_{bb'}\hat{r}_a\hat{r}_{a'}
+ \delta_{ab}\hat{r}_{a'}\hat{r}_{b'}
+ \delta_{a'b'}\hat{r}_{a}\hat{r}_{b})\nonumber \\
&& + C_5 \hat{r}_a\hat{r}_b \hat{r}_{a'}\hat{r}_{b'}\,, 
\end{eqnarray}
where
\begin{eqnarray}
C_1 & = & E_0 + \frac{2}{r}\frac{dE_2}{dr} 
+ \frac{1}{r^2}\frac{d^2E_4}{dr^2} - \frac{1}{r^3}\frac{dE_4}{dr}\,, \\
C_2 & = & -E_0 - \frac{2}{r}\frac{dE_2}{dr} 
+ \frac{1}{r^2}\frac{d^2E_4}{dr^2} - \frac{1}{r^3}\frac{dE_4}{dr}\,,\\
C_3 & = & \frac{d^2 E_2}{dr^2} - \frac{1}{r}\frac{dE_2}{dr}\,, \\
C_4 & = & \frac{1}{r}\frac{d^3 E_4}{dr^3} - \frac{3}{r}\frac{d^2E_4}{dr^2}
+\frac{3}{r^2}\frac{dE_4}{dr}\,, \\
C_5 & = & \frac{d^4 E_4}{dr^4} - \frac{6}{r}\frac{d^3E_4}{dr^3}
+ \frac{15}{r^2}\frac{d^2E_4}{dr^2} - \frac{15}{r^3}\frac{dE_4}{dr}\,.
\end{eqnarray}

The function $E_0$ is the massless scalar two-point function:
\begin{equation}
E_0 = \frac{1}{4\pi^2}\frac{1}{r^2-T^2}\,.
\end{equation} 
The functions $E_2$ and $E_4$ can be evaluated as follows.  From the
formula~\cite{Grad}
\begin{equation}
\int_0^\infty\left[ \frac{\cos x}{x} - \frac{1}{x(1+x)}\right]\,dx = 
- \gamma\,,
\end{equation}
where $\gamma = 0.577...$ is Euler's constant, one finds
\begin{equation}
\lim_{x_0\to 0}\left[ \int_{x_0}^\infty \frac{\cos x}{x}\,dx
+ \log x_0\right] = -\gamma\,.
\end{equation}
This formula and $\int_0^{\pi/2}(\sin x/x)dx = \pi/2$ 
can be used to obtain the following result:
\begin{equation}
\int_\alpha^\infty \frac{dk}{k}e^{ikx}
=  K  -\log \alpha x  + s(x,\alpha) \label{Klog}
\end{equation}
with $K \equiv -\gamma + i\pi/2$, where the error term is 
\begin{equation}
s(x,\alpha) =  -\sum_{n=1}^\infty \frac{(i\alpha x)^n}{n\cdot n!}\,,
\end{equation}
which tends to zero as $\alpha\to 0$.  We neglect the error term from now on.
By integrating by parts and using (\ref{Klog}) with $s(x,\alpha)=0$
we find
\begin{eqnarray}
\int_{\alpha}^\infty \frac{dk}{k^2} e^{ikx}
& = & \frac{1}{\alpha} + ix(K + 1 - \log \alpha x)\,, \\
\int_{\alpha}^\infty \frac{dk}{k^4} e^{ikx}
& = & \frac{1}{3\alpha^3} + \frac{ix}{2\alpha^2} - \frac{x^2}{2\alpha}
- \frac{ix^3}{6} \left( K + \frac{11}{6} - \log \alpha x\right)\,. 
\end{eqnarray}
By using these formulas in (\ref{En}) we find
\begin{eqnarray}
E_2 & = & \frac{1}{4\pi^2}\left[
K+1 - \frac{1}{2}\log \alpha^2(T^2-r^2) - \frac{T}{2r}\log\frac{T+r}{T-r}
\right]\,,\\
E_4 & = & \frac{1}{4\pi^2}\left\{ \frac{1}{2\alpha^2}
+ \frac{iT}{\alpha} -\frac{1}{12}(3T^2+r^2)
\left[ 2K + \frac{11}{3}-\log\alpha^2(T^2-r^2)\right]\right. \nonumber \\
&& \left. + \frac{1}{12}\left(\frac{T^3}{r}+3rT\right)
\log\frac{T+r}{T-r}\right\}\,.
\end{eqnarray}
The functions $C_i$, $i=1,2,3,4,5$, can be obtained by a straightforward
calculation.
By noting that $g_{ab}\to\delta_{ab}$, $g_{ab'} \to \delta_{ab'}$,
$g_{a'b'}\to \delta_{a'b'}$, $N_a \to \hat{r}_a$ and
$N_{a'} \to -\hat{r}_{a'}$ in the flat-space limit,
we can identify the flat-space counterparts 
of the functions $g^{(1)}$, $g^{(2)}$ and $g^{(3)}$ in section 4.  We find
that these functions are indeed given by the right-hand sides of
(\ref{flat1}), (\ref{flat2}) and (\ref{flat3}).

\section*{Appendix B.  Three coordinate systems for de Sitter spacetime}

\setcounter{equation}{0}
\renewcommand{\theequation}{B\arabic{equation}}

The reference for this Appendix is Ref.\ \cite{HawkEllis}.
de Sitter spacetime is the hypersurface in 5-dimensional
Minkowski spacetime with cartesian coordinates
$(T,W,X,Y,Z)$ determined by
\begin{equation}
X^{2}+Y^{2}+Z^{2}+W^{2}-T^{2} = 1/H^2\,.  \label{des}
\end{equation}
The spatially flat coordinate system can be obtained by letting 
\begin{eqnarray}
t & = & \frac{1}{H}\log \left[H(W+T)\right]\,, \label{flt1}\\
x & = & \frac{X}{H(W+T)}\,,\label{flt2}\\
y & = & \frac{Y}{H(W+T)}\,,\label{flt3}\\
z & = & \frac{Z}{H(W+T)}\,. \label{flt4}
\end{eqnarray}
The metric in this coordinate system is given by
(\ref{flat}).
The large-coordinate-distance limit $(x^2+y^2+z^2)^{1/2} \to \infty$ with
$t$ fixed corresponds to
\begin{equation}
X^{2}+Y^{2}+Z^{2}\rightarrow\infty\,\,{\rm with}\,\, W+T = {\rm const}\,.
\end{equation}
Using (\ref{des}), we find that $W-T \to -\infty$ (and $T\to +\infty$,
$W \to -\infty$) in this limit.  

The coordinate system used in the 
main part of this paper, that with $S^3$ spatial
sections, is defined by
\begin{eqnarray}
T & = & H^{-1}\tan \tau\,, \label{HT}\\
W & = & H^{-1}\sec\tau \cos\chi\,,\\
X & = & H^{-1}\sec\tau\sin\chi\sin\theta\cos\varphi\,,\\
Y & = & H^{-1}\sec\tau \sin\chi \sin\theta\sin\varphi\,,\\
Z & = & H^{-1}\sec\tau\sin\chi\cos\theta\,. 
\end{eqnarray}
In the large-distance limit in spatially flat coordinate system, we have 
$T = H^{-1}\tan\tau \to +\infty$,  
but $W+T = H^{-1}(\sin\tau+\cos\chi)/\cos\tau$ stays
constant.  Hence we have $\tau \to \pi/2$ and $\chi \to \pi$.

The coordinate system with hyperbolic spatial sections is given by
\begin{eqnarray}
T & = & H^{-1}\sinh\eta\cosh\zeta\,, \\
W & = & H^{-1}\cosh\eta\,, \\
X & = & H^{-1}\sinh\eta\sinh\zeta\sin\theta\cos\varphi\,, \\
Y & = & H^{-1}\sinh\eta\sinh\zeta\sin\theta\sin\varphi\,, \\ 
Z & = & H^{-1}\sinh\eta\sinh\zeta\cos\theta\, 
\end{eqnarray}
with the metric (\ref{hyperb}).  Since $W = H^{-1} \cosh\eta \geq H^{-1}$, 
the large-distance limit of the spatially flat coordinate system
with $W \to -\infty$ cannot be studied with the hyperbolic coordinate system. 
In the limit $\zeta \to \infty$ we have
$T\to +\infty$ and $X^2+Y^2+Z^2 \to \infty$ with $W$ fixed.  This limit
corresponds to the limit 
$t\to \infty$ with $(x^2+y^2+z^2)^{1/2}\to H^{-1}$ in the spatially flat
coordinate system, and to the limit $\tau\to \pi/2$ with $\chi\to \pi/2$ in
the coordinate system with $S^3$ spatial sections.

\end{document}